\documentclass[12pt]{article}
\usepackage{graphicx, epsfig}
\usepackage{amsmath, amssymb}

\makeatletter
\renewcommand\section{\@startsection {section}{1}{\z@}%
                                   {-3.5ex \@plus -1ex \@minus -.2ex}
                                   {2.3ex \@plus.2ex}%
                                   {\normalfont\large\bfseries}}
\renewcommand\subsection{\@startsection{subsection}{2}{\z@}%
                                     {-3.25ex\@plus -1ex \@minus -.2ex}%
                                     {1.5ex \@plus .2ex}%
                                     {\normalfont\bfseries}}

\makeatother

\newcommand{\bea}{\begin{eqnarray}}
\newcommand{\eea}{\end{eqnarray}}
\newcommand{\be}{\begin{equation}}
\newcommand{\ee}{\end{equation}}

\newcommand{\noi}{\noindent}

\newcommand{\Tr}{\mathrm{Tr}}

\newcommand{\re}{\mathrm{Re}}

\newcommand{\CN}{\mathcal{N}}

\begin{document}

\begin{titlepage}

\begin{center}

\hfill ITFA-08-08

\vskip 2 cm {\Large \bf On the space of elliptic genera} \vskip 1.25 cm
{Jan Manschot}\footnote{Current address: New High Energy Theory Center, Rutgers University, Piscataway, NJ 08854-8019, USA}\\
{\vskip 0.5cm  Institute for Theoretical Physics\\ University of Amsterdam\\
Valckenierstraat 65\\
1018 XE, Amsterdam\\
The Netherlands\\}

\pagestyle{plain}
\end{center}

\vskip 2 cm

\begin{abstract}
\baselineskip=18pt
\noi 
Invariance under modular transformations and spectral flow restrict the
possible spectra of superconformal field theories (SCFT). This paper presents a technique to calculate the number of constraints on the
polar spectra of $\mathcal{N}=(2,2)$ and $\CN=(4,0)$ SCFT's by analyzing their elliptic
genera. The polar spectrum corresponds to the
principal part of a Laurent expansion derived from the elliptic
genus. From the point of view of the AdS$_3$/CFT$_2$ correspondence, these are the states
which lie below the cosmic censorship bound in classical gravity. The dimension
of the space of elliptic genera is obtained as the number of
coefficients of the principal part minus the number of constraints. As
an additional illustration of the technique, the constraints on the
spectrum of $\CN=4$ topologically twisted Yang-Mills on $\mathbb{CP}^2$ are
discussed.    
\end{abstract}
 
\end{titlepage}

\baselineskip=19pt

\tableofcontents


\section{Introduction}

Because two-dimensional conformal field theories (CFT) are
diffeomorphism invariant, partition functions of CFT's
are modular invariant \cite{Cardy:1986ie}. This invariance has major
implications for the spectrum of the CFT. An example is the asymptotic
growth of degeneracies which is given by the celebrated Cardy formula. Modular
invariance also imposes constraints on the part of the spectrum with a 
small number of degeneracies (compared to the regime of validity of
the Cardy formula). Examples of such constraints are 
charge sum rules in superconformal field theories (SCFT)
\cite{Aharony:1993zu}, and constraints on topological quantities of the
target manifold when a sigma model is considered \cite{Witten:1986bf, Eguchi:1988vra,
  gritsenko-991}. In these cases, the constraints are derived by an
analysis of the elliptic genus, which counts states with an
alternating sign dependent on the fermion number. The elliptic genus,
being a supersymmetric index, enumerates BPS states and is well-protected
against continuous changes of parameters which preserve supersymmetry. 
 This makes the elliptic genus an important tool
for the microscopic account of, for example, the entropy of D1-D5 brane systems
\cite{Strominger:1996sh} and M-theory black holes \cite{Maldacena:1997de}. 

The present paper continues the study of implications on SCFT
spectra by an analysis of the $\CN=(2,2)$ and $\CN=(4,0)$ elliptic genus. The symmetries of the
elliptic genus can be derived from modular invariance and the spectral
flow symmetry of the SCFT. It is shown that these symmetries 
impose constraints on the polar spectrum, \footnote{When the words ``degeneracy'' or ``spectrum''
  are used in the following, counting of the states with $(-1)^F$ is 
 implicitly assumed.} whose degeneracies are generically small. Section \ref{sec:egandjf} explains
the notion of ``polar spectrum''. In the context of the
AdS$_3$/CFT$_2$ correspondence \cite{Maldacena:1997re}, the polar
spectrum is that part of the spectrum which classically lies below the
cosmic censorship bound of AdS$_3$-gravity \cite{Cvetic:1998xh}. This part of
the spectrum is of particular importance in
Ref. \cite{Dijkgraaf:2000fq}, where a gravity interpretation is given
to the partition function of the boundary SCFT using the method of
images. Ref. \cite{Manschot:2007ha} shows that not every set of
polar degeneracies can be completed to
a full partition function by this method. Certain spectra are thus easily excluded as CFT
 spectra, based on their inconsistency with the required
 symmetries. Gaberdiel {\it et al.} apply in Ref. \cite{gaberdiel:2008} the presence of constraints on $\CN=(2,2)$ polar
 spectra, to analyze the consistency of ``pure''
 $\mathcal{N}=2$ supergravity with the SCFT symmetries. Of course, the existence of a partition function with the
 right properties does not imply the existence of a CFT. In case a CFT does exist and constraints are present, only a subset
 of the polar degeneracies  need to be specified to determine the
 complete partition function. For large central charges of $\CN=(2,2)$
 and $\CN=(4,0)$ SCFT's, the ratio of the number of  polar
 degeneracies and the number of constraints is shown to grow linearly
 with the central charge.

 Consistency of a given polar spectrum with the
 symmetries can be tested by the construction of a function (a vector-valued
 cusp form \footnote{Cusp forms are holomorphic modular forms which
   vanish at   the cusps, which are the points
   $i\infty\cup\mathbb{Q}$.}), which is determined by the polar degeneracies. When this
 function is non-vanishing, no partition function with
 the right properties exists. This is explained in 
 Ref. \cite{Manschot:2007ha}, which relied on methods described in
 \cite{niebur:1974}. The number of constraints on the polar spectrum is equal to the
 dimension of the space of appropriate vector-valued cusp forms. 
The present paper describes a technique, following the original work of Ref.
\cite{skoruppa:1984}, to calculate the dimension of
the space of such cusp forms. By a more straightforward approach, simpler
expressions for the dimension formulas for vector-valued cusp forms
are derived compared to the formulas presented in
\cite{skoruppa:1984}.  

As explained in Section \ref{sec:egandjf}, an $\CN=(2,2)$ elliptic
genus is an element of the space of weak Jacobi forms $\tilde
J_{0,m}$ with an integral Fourier expansion. We will often refer to the
space $\tilde J_{0,m}$ as the space of elliptic genera, although not
even all functions with an integral expansion in $\tilde J_{0,m}$
appear as SCFT elliptic genera. Having obtained the dimension
of the space of vector-valued cusp forms, one can calculate the dimension of the
space of elliptic genera straightforwardly. The final expression for
$\dim \tilde J_{0,m}$ (Eq. (\ref{eq:j(m)})) equals the dimension
formula given in Ref. \cite{eichler}. However, the derivation here is qualitatively
 different. The technique  described here has the advantage that it
 is more generally valid than the technique of \cite{eichler}, in
 physical applications where modular forms appear.  Section \ref{sec:constraints04} applies
 the technique to elliptic genera of  $\mathcal{N}=(4,0)$ SCFT's,
 which are relevant for the  microscopic explanation of the entropy of
 M-theory black holes  \cite{Maldacena:1997de, Gaiotto:2006wm,
 Kraus:2006nb, deBoer:2006vg}.  As an additional illustration,
the dimension of the space of weakly holomorphic functions is calculated, which satisfy the expected transformation properties of the 
partition functions of topologically twisted $\CN=4$ supersymmetric $SU(N)$
Yang-Mills. The constraints are more restrictive in this case than for the $\CN=(4,0)$ SCFT's arising in M-theory. Another
application is the calculation of the dimension of spaces of 
characters in rational conformal field  theories, which is not
 attempted here.

Section \ref{sec:egandjf} starts with a review of the $\mathcal{N}=(2,2)$
elliptic genus and its connection with weak Jacobi forms. The number of
independent constraints is shown to equal the dimension of a certain
space of vector-valued cusp forms in Section \ref{sec:contcusp}. The dimension of
the space of cusp forms is calculated. The number of polar
degeneracies minus this dimension gives the dimension of the space of
elliptic genera. The application to $\CN=(4,0)$ elliptic genera and 
$\CN=4$ Yang-Mills theory is discussed in Section \ref{sec:constraints04}.

\section{$\CN=(2,2)$ elliptic genera and Jacobi forms}
\setcounter{equation}{0}
\label{sec:egandjf}
The elliptic genus is defined as the trace over the Ramond-Ramond
sector of an SCFT,
\be
\label{eq:ellipticgenus}
Z(\tau,z)=\mathrm{Tr}_{\mathrm{RR}}(-)^Fy^{J_0}q^{L_0-c_L/24}\bar q^{\bar L_0-c_R/24},
\ee
where $q=e(\tau)$, \footnote{We use the convention $e(x)=e^{2\pi i
    x}$.} $\tau\in \mathcal{H}$ where $\mathcal{H}$ is the upper-half
plane, and $y=e(z)$, $z\in \mathbb{C}$. The insertion of $(-)^F$ (with 
$F=\frac{1}{2}(J_0-\bar J_0)$ being the fermion number) projects the
trace on states which preserve the supersymmetries in the 
right-moving sector. Since these are the right-moving ground states,
the trace is independent of $\bar \tau$. 

Modular invariance of the CFT implies that $Z(\tau,z)$ transforms under
a modular transformation as \cite{Kawai:1993jk}
\be
\label{eq:modinv}
Z\left( \gamma(\tau),
\frac{z}{j(\gamma,\tau)}\right)=e\left(\frac{m c z^2}{j(\gamma,\tau)}\right)Z(\tau,z), \qquad \gamma=\left( \begin{array}{cc} a & b \\ c & d
\end{array} \right)\in \Gamma, \qquad m=\frac{c_L}{6},\\
\ee
where $j(\gamma,\tau)=c\tau+d$, and $\Gamma$ is an abbreviation for
$SL(2,\mathbb{Z})$. When the SCFT is a sigma model with $d$-dimensional target
space, then $m=\frac{d}{4}$. Restricting to specific
    values of $z$ in this case, gives several topological quantities of the target
    manifold \cite{Eguchi:1988vra}. 

Spectral flow is a symmetry of the superconformal algebra, which relates
states with different periodicities of the fermions. States in the
Ramond sector can for example be mapped to states in the Neveu-Schwarz sector. The
requirement that the spectrum satisfies this symmetry,
implies that $Z(\tau,z)$ is quasi-periodic as a function of $z$
\cite{Kawai:1993jk}. The quasi-periodicity is given by 
\be
\label{eq:specflow}
Z\left(\tau, z+\lambda \tau
+\mu\right)=(-1)^{2m(\lambda+\mu)}e\left(-m (\lambda^2 \tau + 2
\lambda z )\right)Z(\tau,z), \qquad (\lambda, \,\, \mu) \in
\mathbb{Z}^2.
\ee
 This paper considers elliptic genera
with integer $m$; the results are easily generalized to the case of
non-integer $m$. The symmetries (\ref{eq:modinv}) and 
(\ref{eq:specflow}) determine that the elliptic genus is a (weak)
Jacobi form of weight $0$ and index $m$ \cite{eichler}. The adjective ``weak'' will
be explained below. The space of weak Jacobi forms of weight 0 and
index $m$ is denoted by $\tilde J_{0,m}$. 

Many aspects of Jacobi forms can be understood via the connection
between Jacobi forms and vector-valued modular forms of half-integer
weight. This connection is briefly outlined here; additional details and
proofs can be found in Ref. \cite{eichler}. A weak Jacobi form
$\phi(\tau,z)$ has a Fourier expansion in terms of the integer coefficients $c(n,l)$
\be
\label{eq:fourierjacobi}
\phi(\tau, z)=\sum_{n\geq 0, l\in \mathbb{Z}}c(n,l) q^n y^l.
\ee
\noi Spectral flow (\ref{eq:specflow}) determines $c(n,l)$ to be a function of only 
$4mn-l^2$ and the residue $l\mod 2m$. The adjective ``weak'' in ``weak Jacobi form'' is
used when $c(n,l)$ is non-zero for $4mn-l^2\geq -m^2$, as opposed to
$\geq 0$ for a Jacobi form. From spectral flow symmetry, or
equivalently quasi-periodicity, one can deduce that $\phi(\tau,
z)$ admits a decomposition into a set of functions
$h_\mu(\tau)$ and theta functions $\theta_{m,\mu}(\tau,z)$ with
$\mu\in \mathbb{Z}/2m\mathbb{Z}$. In terms
of these functions, $\phi(\tau, z)$ is given  by 
\begin{equation}
\label{eq:thetadecomposition}
\phi(\tau, z)=\sum_{\mu \mod 2m} h_\mu(\tau) \theta_{m,\mu}(\tau,z).
\end{equation}
The functions $h_\mu(\tau)$ and $\theta_{m,\mu}(\tau,z)$ are given
explicitly by 
\begin{equation}
\label{eq:fourierjacobidecomp}
h_\mu(\tau)=\sum_{n=-\mu^2\mod 4m} c_\mu(n) q^{n/4m}, \qquad
\theta_{m,\mu}(\tau,z)=\sum_{{l\in \mathbb{Z} \atop l=\mu \mod 2m}}q^{l^2/4m}y^l,
\end{equation}
\noi where $c_\mu(n)=(-1)^{2ml}c(\frac{n+l^2}{4m},l)$, and $l=\mu \mod
2m$. The domain of $\mu$ in Eq. (\ref{eq:thetadecomposition}) is
taken to be $\left[-m+1, m \right]$.  

All the information concerning the Fourier coefficients of $\phi(\tau,z)$ is
thus captured in $h_\mu(\tau)$. The space $\tilde J_{0,m}$ is
therefore isomorphic to the space of appropriate vector-valued modular
forms. The precise transformation properties of the vector
$h_\mu(\tau)$ are described later in this section. The adjective
``weak'' implies that the Laurent expansion of $h_\mu(\tau)$ may admit
a principal part, which corresponds to the terms with negative
exponents, $-m/4\leq n/4m<0$.  The negative exponents lead to a pole
of $h_\mu(\tau)$ in the limit $\tau \to i \infty$. By
$\Gamma$-transformations, the limit $\tau\to i\infty$ of the vector
$h_\mu(\tau)$ is equivalent to $\tau \to \mathbb{Q}$. Modular forms which are only 
meromorphic for $\tau \to i\infty \cup \mathbb{Q}$ are called ``weakly holomorphic''. 
The poles of the $h_\mu(\tau)$ are weak enough to be
canceled by the theta functions, such that $\phi(\tau,z)$ is an
analytic function. The coefficients $c_\mu(n)$, $n<0$, are referred to as ``polar
coefficients''. We denote the space of polar coefficients for a given 
index $m$ by $P(m)$; the dimension of $P(m)$ is $p(m)$. These
coefficients correspond to the terms with $4mn-l^2<0$ in expansion
(\ref{eq:fourierjacobi}). 

The polar spectrum is defined as the set of
states, which is counted by the principal part of $h_\mu(\tau)$. From the point of view of the
SCFT, the condition $4mn-l^2<0$ corresponds to states, with eigenvalues
$\frac{2}{3}c_L(L_0-\frac{c_L}{24})-J_0^2<0$. The notion of ``polar spectrum'' is more
generally valid in CFT, for example the polar spectrum of a bosonic CFT is the set of states
with $L_0-\frac{c_L}{24}<0$. In AdS$_3$, a black hole with such quantum numbers would classically lead to a
naked singularity. Therefore, the polar states lie below the cosmic
censorship bound from the viewpoint of AdS$_3$-gravity
\cite{Dijkgraaf:2000fq, Cvetic:1998xh}. These states are interpreted as
excitations of thermal AdS$_3$, whereas the non-polar states are
mostly contributed to black hole geometries \cite{Dijkgraaf:2000fq, Witten:2007kt}.

The functions $h_\mu(\tau)$ and $\theta_{m,\mu}(\tau,z)$ transform under $\Gamma$ with 
half-integral weight, which requires the appearance of non-trivial unitary factors
in modular transformations \cite{koblitz}. Therefore, a double sheeted
cover of $\Gamma$, the metaplectic group $\tilde \Gamma$, is first
introduced. An element $\tilde \gamma \in \tilde \Gamma$ is represented by 
\be
\label{eq:metael}
\tilde \gamma= \left(\gamma,\epsilon \sqrt{j(\gamma,\tau)}\right), \qquad \gamma \in \Gamma,\qquad
\epsilon=\pm 1.
\ee
The square root of $j(\gamma,\tau)$ is defined unambiguously by
requiring that the argument of a complex number $z$ lies in the domain
$(-\pi, \pi]$. The product of two elements is defined by
\be
\left(\gamma,\epsilon \sqrt{j(\gamma,\tau)}\right)\cdot
\left(\gamma',\epsilon'
  \sqrt{j(\gamma',\tau)}\right)=\left(\gamma\gamma',\epsilon\epsilon'\sqrt{j(\gamma,\gamma'(\tau))}\sqrt{j(\gamma',\tau)}\right). 
\ee
We define the slash operator $|_w\tilde \gamma$ on a modular form $f(\tau)$ of (possibly
half-integer) weight $w$, by
\be
f|_{w}\tilde \gamma=\epsilon^{-2w}j(\gamma,\tau)^{-w}f\left(\gamma(\tau)\right),
\ee
and the slash operator $|_{k,m}\tilde \gamma$ on Jacobi forms $\phi(\tau,z)$ of weight $k$
and index $m$, by
\be
\phi|_{k,m}\tilde \gamma=\epsilon^{-2k}j(\gamma,\tau)^{-k}e\left(-\frac{mcz^2}{j(\gamma,\tau)}\right)\phi\left(\gamma(\tau),\frac{z}{j(\gamma,\tau)}\right).
\ee

The set of theta functions $\theta_{m,\mu}$ transforms as a
vector-valued Jacobi form under transformations $\tilde \gamma\in \tilde \Gamma$:
\be
\label{eq:thetatransgamma}
\left(\begin{array}{c}\theta_{m,-m+1}|_{\frac{1}{2},m}\tilde \gamma \\
    \theta_{m,-m+2}|_{\frac{1}{2},m}\tilde \gamma \\ \dots \\
  \dots \\ \theta_{m,m}|_{\frac{1}{2},m}\tilde \gamma \end{array}
\right)=\mathbf{M}^\mathrm{T}_{2m}(\tilde \gamma) \left(\begin{array}{c}\theta_{m,-m+1} \\ \theta_{m,-m+2} \\ \dots \\
  \dots \\ \theta_{m,m} \end{array} \right),
\ee
where $\mathbf{M}_{2m}(\tilde \gamma)$ is a $2m\times 2m$ matrix. The matrix $\mathbf{M}_{2m}(\tilde \gamma)$ 
appears transposed in (\ref{eq:thetatransgamma}) for notational clearity in the rest of the text. 

Generators of $\tilde \Gamma$ are $S=\left(\left(\begin{array}{cc}0
  & -1 \\ 1 & 0 \end{array}\right), \sqrt{\tau} \right)$ and $T=\left(\left(\begin{array}{cc}1
  & 1 \\ 0 & 1 \end{array}\right), 1 \right)$. The transformations of $\theta_{m,\mu}(\tau,z)$ under $S$ and
  $T$ are given by  
\begin{eqnarray}
\label{eq:thetatransform}
&\theta_{m,\mu}\vert_{\frac{1}{2},m}\, S&=\frac{1}{\sqrt{2mi}} \sum_{\nu \mod 2m}
e\left(-\frac{\mu \nu}{2m}\right) \theta_{m,\nu}(\tau,z), \\
&\theta_{m,\mu}\vert_{\frac{1}{2},m}\, T&=e\left(\frac{\mu^2}{4m}\right)\theta_{m,\mu}(\tau,z). \nonumber
\end{eqnarray}
These transformations are in principle sufficient to deduce
$\mathbf{M}_{2m}(\tilde \gamma)$ for general $\tilde \gamma\in \tilde
\Gamma$. Closed expressions are also known \cite{skoruppa:1984}. For elements
$\tilde \gamma \in \tilde \Gamma$ which lie in the congruence subgroup
$\Gamma(4m)^* \in \tilde \Gamma$, $\mathbf{M}_{2m}(\tilde \gamma)$ is the identity
matrix. The group $\Gamma(4m)^*$ is defined by 
\be
\label{eq:4m*}
\Gamma(4m)^*=\left\{\left.\tilde \gamma=\left(\gamma, \left(\frac{c}{d}\right) \varepsilon_d^{-1}
j(\gamma,\tau)^{\frac{1}{2}}\right)\right| \gamma \in \Gamma(4m) \right\},
\ee
where $\Gamma(N)\in \Gamma$ is the principal congruence subgroup
\be
\Gamma(N)=\left\{ \left(\begin{array}{cc}a & b \\ c & d \end{array}
  \right)\in \Gamma, \left(\begin{array}{cc}a & b \\ c & d \end{array}
  \right)= \left(\begin{array}{cc}1 & 0 \\ 0 & 1 \end{array}
  \right)\mod N \right\}.
\ee 
The groups $\Gamma(4m)^*$ and $\Gamma(N)$ are normal subgroups of
respectively $\tilde \Gamma$ and $\Gamma$.
In (\ref{eq:4m*}), $\left(\frac{c}{d} \right)$ is the extended Legendre symbol \cite{koblitz} and
$\varepsilon_d=\sqrt{\left(\frac{-1}{d}\right)}$,
\be
\label{eq:epsd}
\varepsilon_d= \left\{\begin{array}{cc} 1, & d=1 \mod 4, \\  i, & d=3 \mod 4.
\end{array}\right.
\ee
Eq. (\ref{eq:4m*}) gives an explicit expression for $\epsilon$ in
Eq. (\ref{eq:metael}). This expression is derived from the
transformation properties of the theta function
$\Theta(\tau)=\sum_{n\in \mathbb{Z}}q^{n^2}$ under $\Gamma_0(4)$, and is therefore consistent
with the transformations of half-integer weight forms
\cite{koblitz}. Since this expression for $\epsilon$ takes values in
$(\pm 1, \pm i)$, $\Theta(\tau)$ actually transforms under a
four-sheeted cover of $\Gamma$. Using the transformations in 
Eq. (\ref{eq:thetatransform}), $\theta_{m,\mu}(\tau,z)$  can be shown 
to transform diagonally under $\Gamma(4m)^*$, \footnote{In the following, the
  tilde is omitted from elements $\tilde \gamma \in \tilde \Gamma$.}  
\be
\theta_{m,\mu}|_{\frac{1}{2},m}\gamma=\theta_{m,\mu}(\tau,z) \,\,, \qquad \gamma\in \Gamma(4m)^*.
\ee
Note that $\theta_{m,\mu}(\tau, z)$ is not multiplied by an additional
unitary pre-factor; $\mathbf{M}_{2m}(\gamma)$ is thus indeed the
identity matrix for $\gamma \in \Gamma(4m)^*$. More general
transformations, acting diagonally on $\theta_{m,\mu}(\tau, 
z)$ but with non-trivial unitary pre-factor, form a larger congruence subgroup. 

From the above considerations, we deduce that the matrices $\mathbf{M}_{2m}(\gamma)$ form a $2m$-dimensional
representation of the finite group $\tilde \Gamma/\Gamma(4m)^*$. Since the
generators of $\tilde \Gamma$, $S$ and $T$, are both represented by unitary
matrices $\mathbf{M}_{2m}(S)$ and $\mathbf{M}_{2m}(T)$, the representation
is unitary.

The transformations of the theta functions combined with those of $\phi(\tau,z)$
(given by (\ref{eq:modinv}) in terms of $Z(\tau,z)$) determine that the functions $h_\mu(\tau)$ transform as a
vector-valued modular form with weight $-\frac{1}{2}$, and conjugately to the transformations of $\theta_{m,\mu}$,
\be
\left(\begin{array}{c}h_{-m+1}|_{-\frac{1}{2}}\gamma \\
    h_{-m+2}|_{-\frac{1}{2}}\gamma \\ \dots \\
  \dots \\ h_{m}|_{-\frac{1}{2}}\gamma \end{array}
\right)=\mathbf{M}^{-1}_{2m}(\gamma) \left(\begin{array}{c}h_{-m+1} \\ h_{-m+2} \\ \dots \\
  \dots \\ h_{m} \end{array} \right),
\ee
for $\gamma \in \tilde \Gamma$. Since the representation is unitary,
$\mathbf{M}^{-1}_{2m}(\gamma)=\overline{\mathbf{M}_{2m}(\gamma)}$.

The modular forms $h_\mu(\tau)$, $\mu\in \mathbb{Z}/2m\mathbb{Z}$, are
not all linearly independent forms. This is a consequence of the fact
that $(-1,1)$ acts non-trivially on $\theta_{m,\mu}(\tau, z)$ but
leaves $h_\mu(\tau)$ invariant. The equality $\phi\vert_{0,m}(-1,1)=\phi$, which is
equivalent to
\be
\label{eq:z-z}
\phi(\tau,z)=\phi(\tau,-z),
\ee
requires then that $h_{-\mu}(\tau)=h_{\mu}(\tau)$. The modular
transformations of the vector-valued modular form $h_\mu(\tau)$ are
therefore adequately described by a vector of
length $m+1$, and an $(m+1)\times (m+1)$ matrix $\overline{\mathbf{M}(\gamma)}$. This gives rise to an
$(m+1)$-dimensional representation of the finite group $\tilde
\Gamma/\Gamma(4m)^*$. The domain of $\mu$ in this representation is taken to be
$\left[0,m\right]$.  

\section{Cusp forms as constraints on $\CN=(2,2)$ spectra}
\setcounter{equation}{0}
\label{sec:contcusp}

This section calculates the number of independent constraints on the polar
spectrum, which is the number of constraints on the polar coefficients
$c_\mu(n)$, $n<0$, of a Jacobi form. This result, given in
(\ref{eq:dimcusp}), allows us to determine the dimension $\tilde
J_{0,m}$ or equivalently the space of elliptic genera in (\ref{eq:j(m)}).

Before considering the constraints, first is shown that the weakly holomorphic, negative
weight modular form $h_\mu(\tau)$ in (\ref{eq:thetadecomposition}) and
(\ref{eq:fourierjacobidecomp}) is uniquely determined by its polar
coefficients. One would namely find a contradiction when two or more
such forms exist, having the same principal part but different regular
part. The difference of two such forms would be a holomorphic, negative
weight modular form of $\Gamma(4m)^*$. Since such holomorphic forms with
negative weight do not exist, the polar coefficients uniquely
determine the weakly holomorphic modular form. The upper bound on the dimension of weakly holomorphic,
negative weight modular forms, with a given maximum order of the pole
at $\tau\to i\infty$, is therefore given by the number of polar terms $p(m)$.


One encounters the presence of constraints when one attempts to
find a negative weight modular form with a prescribed set of polar
coefficients \cite{niebur:1974, Manschot:2007ha}. Naively, a sum over
$\Gamma/\Gamma_\infty$, which is known as a Poincar\'e series,
completes a function which is not modular covariant to a modular
covariant object. However, Refs. \cite{niebur:1974, Manschot:2007ha} 
explain that a sum over $\Gamma/\Gamma_\infty$ does not complete every
possible choice of polar terms to a modular form or Jacobi
form. Ref. \cite{Manschot:2007ha} aims to construct a
Jacobi form by a Poincar\'e series on the 
principal part of $h_\mu(\tau)$. For a general choice of $c_\mu(n)$,
$n<0$, the constructed function $\phi(\tau,z)$ does not transform
as a form in $\tilde J_{0,m}$. Instead, $\phi(\tau,z)$ transforms with
an anomalous shift under $\Gamma$-transformations  
\be
\label{eq:transgenpol}
\left.\phi\right|_{0,m}\gamma-\phi=-\frac{1}{\Gamma(3/2)}\sum_{\mu\!\!
  \mod 2m}\theta_{m,\mu}(\tau, z)\int_{\gamma^{-1}(\infty)}^{-i\infty}
\overline{g_\mu(t)}(\bar t-\tau)^{\frac{1}{2}}d\bar t ,
\ee
where $\Gamma(x)$ is the Gamma-function. The vector $g_\mu(\tau)$ is a
vector-valued cusp form of weight $2\frac{1}{2}$ and it transforms conjugately to $h_\mu(\tau)$, thus with the $(m+1)\times
(m+1)$ matrices $\mathbf{M}(\gamma)$. We refer to the space of these
vector-valued cusp forms as $S_{2\frac{1}{2},\mathbf{M}}(\Gamma(4m)^*)$;
$S_{2\frac{1}{2},\mathbf{M}}(\Gamma(4m)^*)$ has the argument
  $\Gamma(4m)^*$ since the matrices $\mathbf{M}(\gamma)$ form a representation of
  $\tilde \Gamma/\Gamma(4m)^*$. The right hand side does not vanish
unless $g_\mu(\tau)=0$ \cite{niebur:1974}. Moreover, independent
vector-valued forms $g_\mu(\tau)$ lead to independent vector-valued
functions of $\tau$ \cite{knopp:1974} after the integration over $\bar
t$ in (\ref{eq:transgenpol}).

The cusp form $g_\mu(\tau)$ is a Poincar\'e series determined
by the coefficients $c_\mu(n)$, $n<0$ \cite{niebur:1974, Manschot:2007ha}. Vanishing of $g_\mu(\tau)$ is
established for a proper choice of $c_\mu(n)$, $n<0$, \footnote{In the
  case of bosonic pure gravity \cite{Witten:2007kt}, the 
  corresponding cusp form would be a weight two cusp form of
  $\Gamma$. Since these do not exist, no constraints by modularity are
  imposed on the polar spectrum, which is consistent with
  \cite{Witten:2007kt}.} namely a choice which corresponds to the
coefficients of a weak Jacobi form. To learn whether a given polar
spectrum is consistent with the space of potential elliptic genera, one can 
check whether the corresponding cusp form vanishes or not. Since the Poincar\'e
series span the space of cusp forms \cite{sarnak:1990} and the
integration leads to independent functions of $\tau$, the number of
independent constraints is equal to the dimension of the space of vector-valued cusp
forms. More details can be found in \cite{Manschot:2007ha}. In the context of scalar modular forms,
Ref. \cite{niebur:1974} established a space of cusp forms as an 
obstruction to the construction of non-positive weight modular forms
with singularities.

The space of cusp forms forms thus an obstruction to the construction of
Jacobi forms, and imposes restrictions on the choice of polar
coefficients. The space $P(m)$ reduced by the constraints is denoted by
$P_\mathrm{c}(m)$. Since a specific choice of polar
coefficients which lies in $P_\mathrm{c}(m)$ corresponds to a unique form 
in $\tilde J_{0,m}$, $\dim P_\mathrm{c}(m)$ is equal to $\dim \tilde
J_{0,m}$. This dimension can thus be calculated by
\be
\label{eq:j=p-dim}
\dim \tilde J_{0,m}=p(m)-\mathrm{dim}\, S_{2\frac{1}{2},\mathbf{M}}(\Gamma(4m)^*).
\ee
The different spaces and the described relations
between them can be nicely summarized as a short exact sequence 
\be
0 \to \tilde J_{0,m} \to P(m) \to S_{2\frac{1}{2},\mathbf{M}}(\Gamma(4m)^*) \to 0,
\ee
where the second arrow maps a given Jacobi form to the set of polar
coefficients $c_\mu(n)$, $n<0$, and the third arrow is the
construction of the vector-valued cusp form from the polar
coefficients by a Poincar\'e series \cite{Manschot:2007ha}.

The obstructions can be viewed as a manifestation
of the Mittag-Leffler problem \cite{griffiths}, which is the problem
of finding a meromorphic section with prescribed singularities of a line bundle $\mathcal{L}$ over a
manifold $X$. The space of obstructions to finding such a section is given by
$H^1(X,\mathcal{O}(\mathcal{L}))$, where $\mathcal{O}(\mathcal{L})$ is
the sheaf of holomorphic sections of $\mathcal{L}$. Since the modular
curve $\mathcal{H}/\Gamma$ is one-dimensional,
$H^1(X,\mathcal{O}(\mathcal{L}))$ is by Serre duality related to
$H^0(X, \mathcal{O}(K \times \mathcal{L^*}))$, with $K$ the canonical
bundle. In the present discussion, sections of $\mathcal{L}$ have
weight $-\frac{1}{2}$ and therefore sections of $\mathcal{L^*}$ have
weight $\frac{1}{2}$. Since holomorphic sections of $K$ are cusp forms
of weight two, this explains the appearance of cusp forms of weight $2\frac{1}{2}$ as obstructions to
the construction of elliptic genera. Ref. \cite{borcherds:1999} generalizes these
considerations to the vector-valued case and proves that the
obstruction space for vector-valued modular forms $h_\mu(\tau)$ is
given by the vector-valued cusp forms $g_\mu(\tau)$. Explicit 
calculations of the dimensions of vector-valued modular forms 
are carried out in \cite{skoruppa:1984}. Such dimension formulas are also
mentioned in \cite{borcherds:1999}, \cite{Eholzer:1994ta} and
\cite{Bantay:2007}. 

The basic ingredients for the calculation of $\dim
S_{2\frac{1}{2},\mathbf{M}}(\Gamma(4m)^*)$ are the
orthogonality relations for irreducible characters of finite groups 
and the Selberg trace formula. The relevant finite group appeared in
Section \ref{sec:egandjf}, namely $\tilde \Gamma/\Gamma(4m)^*$. The
transformation properties of $g_\mu(\tau)$ provide an
$(m+1)$-dimensional representation $\mathbf{M}$ in terms of the matrices
$\mathbf{M}(\gamma)$. We define a character of this representation in the usual
way by
\be
\chi_{\mathbf{M}}(\gamma)=\Tr(\mathbf{M}(\gamma)).
\ee
We label the set of irreducible representations by
$\mathbf{R}_i$. The orthogonality relations for characters of
finite groups read in this case
\be
\frac{1}{\vert \tilde \Gamma/\Gamma(4m)^* \vert}\sum_{\gamma \in \tilde \Gamma/\Gamma(4m)^*}\chi_{\mathbf{R}_i}(\gamma)\overline{\chi_{\mathbf{R}_j}(\gamma)}=\delta_{ij}.
\ee
The multiplicities $m_i$ of the irreducible representations $\mathbf{R}_i$
in $\mathbf{M}$ are given by 
\be
m_i=\frac{1}{\vert \tilde
  \Gamma/\Gamma(4m)^* \vert}\sum_{\gamma \in \tilde
  \Gamma/\Gamma(4m)^*}\chi_{\mathbf{M}}(\gamma)\overline{\chi_{\mathbf{R}_j}(\gamma)}.
\ee
Since $\Gamma(4m)^*$ lies in the kernel of the
representation $\mathbf{M}$, the individual vector elements $g_\mu(\tau)$ lie in the space of weight
$2\frac{1}{2}$ cusp forms of $\Gamma(4m)^*$, which is denoted by
$S_{2\frac{1}{2}}(\Gamma(4m)^*)$. Since $\Gamma(4m)^*$ is a normal
subgroup of $\tilde \Gamma$, the space $S_{2\frac{1}{2}}(\Gamma(4m)^*)$ is
closed under transformations of $\gamma \in \tilde \Gamma$; such
transformations rotate a chosen set of basis elements of $S_{2\frac{1}{2}}(\Gamma(4m)^*)$
among each other. As a consequence, $S_{2\frac{1}{2}}(\Gamma(4m)^*)$
defines a $\mathrm{dim}\, S_{2\frac{1}{2}}(\Gamma(4m)^*)$-dimensional
representation of $\tilde \Gamma/\Gamma(4m)^*$, which can similarly be
decomposed into the irreducible representations. When the
multiplicities of $\mathbf{R}_i$ in $S_{2\frac{1}{2}}(\Gamma(4m)^*)$
are $s_i$, then $\mathrm{dim}\,S_{2\frac{1}{2},\mathbf{M}}(\Gamma(4m)^*) =\sum_i m_i
s_i$. The character of the element $\gamma$ in the $\tilde \Gamma/\Gamma(4m)^*$-representation
$S_{2\frac{1}{2}}(\Gamma(4m)^*)$ is denoted by 
\be
\label{eq:traceS}
\mathrm{Tr}\left[\gamma, S_{2\frac{1}{2}}(\Gamma(4m)^*) \right].
\ee
In terms of the characters, $\mathrm{dim}\,S_{2\frac{1}{2},\mathbf{M}}(\Gamma(4m)^*)$ is now expressed by
\be
\label{eq:dimSM}
\mathrm{dim}\,S_{2\frac{1}{2},\mathbf{M}}(\Gamma(4m)^*)=\frac{1}{\vert \tilde
  \Gamma/\Gamma(4m)^* \vert}\sum_{\gamma \in \tilde
  \Gamma/\Gamma(4m)^*}
\chi_{\mathbf{M}}(\gamma)\overline{\mathrm{Tr}\left[\gamma,
    S_{2\frac{1}{2}}(\Gamma(4m)^*) \right]}. 
\ee

The Selberg trace formula provides a way to determine traces as in Eq. (\ref{eq:traceS}), for
example $\mathrm{Tr}\left[\gamma,
  S_{w}(\Gamma(4m)^*)\right]-\mathrm{Tr}\left[\gamma^{-1},
  M_{2-w}(\Gamma(4m)^*)\right]$ can be calculated. The space
$M_{2-w}(\Gamma(4m)^*)$ is the space of holomorphic modular forms of
$\Gamma(4m)^*$ with weight $2-w$. This is applied by Theorem 5.1 of
Ref. \cite{skoruppa:1984} to calculate the difference of the
dimensions of $S_{w,\mathbf{M}}(\Gamma(4m)^*)$ and
$M_{2-w,\mathbf{\bar M}}(\Gamma(4m)^*)$, where $\mathbf{M}$ is a
representation of $\tilde \Gamma/\Gamma(4m)^*$. It is a sum, of three (generically
fractional) contributions    
\be
\label{eq:dims}
\mathrm{dim}\, S_{w,\mathbf{M}}(\Gamma(4m)^*)
-\mathrm{dim}\,M_{2-w,\mathbf{\bar M}}(\Gamma(4m)^*)
=A_\mathrm{s}+A_\mathrm{e}+A_\mathrm{p}.
\ee
The subscripts ``s'', ``e'' and ``p'' refer respectively to ``scalar'', ``elliptic''
and ``parabolic''. This terminology appears naturally in the derivation of
the Selberg trace formula, see for example \cite{zagier:1976}.   
The three contributions are given  by \cite{skoruppa:1984,Eholzer:1994ta}
\begin{eqnarray}
\label{eq:conttrace}
A_\mathrm{s}&=&\frac{w-1}{12}\chi_{\mathbf{M}}(1), \nonumber \\
A_\mathrm{e}&=&\frac{1}{4}\re\left[e\left(\frac{w}{4}\right)\chi_{\mathbf{M}}\left( S
  \right)\right]+\frac{2}{3\sqrt{3}}\re\left[e\left(\frac{2w+1}{12}\right)\chi_{\mathbf{M}}\left(ST \right)\right],  \\
A_\mathrm{p}&=&-\frac{1}{2}S(\mathbf{M})-\sum_{j=1}^d((\lambda_j)),\nonumber
\end{eqnarray}
where the trace of the identity matrix $\chi_{\mathbf{M}}(1)$ is the
dimension $d$ of the representation $\mathbf{M}$. The numbers $\lambda_j$ are the fractional numbers appearing in
$\chi_\mathbf{M}(T^n)=\sum_j e(\lambda_j n)$. The symbol
$S(\mathbf{M})$ is defined as the number of $\lambda_j$ which
take values in $\mathbb{Z}$. The function $((x))$  is defined by
\be
((x))=x-\frac{\lceil x \rceil+\lfloor x \rfloor}{2}=\left\{\begin{array}{cl} \xi-\frac{1}{2}, & \mathrm{if}\,\,\, x=\xi +
\mathbb{Z}, \,\, 0<\xi<1, \\ 0, & \mathrm{if}\,\,\, x\in \mathbb{Z}. \end{array} \right.
\ee

Eq. (\ref{eq:dims}) provides us $\mathrm{dim}\,S_{2\frac{1}{2},\mathbf{M}}(\Gamma(4m)^*)$,
since holomorphic modular forms with negative weight do not exist and hence
$\mathrm{dim}\,M_{-\frac{1}{2},\mathbf{\bar M}}(\Gamma(4m)^*)
=0$. To calculate the dimension of $S_{2\frac{1}{2},\mathbf{M}}(\Gamma(4m)^*)$,
one needs to evaluate $\chi_{\mathbf{M}}(\gamma)$ for the relevant
$\gamma$'s and substitute in Eq. (\ref{eq:conttrace}). We proceed with
a direct evaluation of $\chi_{\mathbf{M}}(\gamma)$ from the matrices
defined in Section \ref{sec:egandjf}. From Eq. (\ref{eq:z-z}) we
deduced that the $h_\mu(\tau)$ are described by an $(m+1)$-dimensional
vector, thus $d=m+1$ and $\mu=0\dots  m$.  One can express $g_\mu\vert_{2\frac{1}{2}}\,S$ and $g_\mu\vert_{2\frac{1}{2}}\,T$ in terms of $g_\mu(\tau)$
 with $\mu=0\dots m$: 
\begin{eqnarray}
\label{eq:gtransform}
g_\mu\vert_{2\frac{1}{2}}\,S&=&\frac{1}{\sqrt{2mi}}\left(g_0(\tau)-e\left(\frac{\mu}{2}\right)g_m(\tau)\right. \\
&&+\sum_{\nu=1}^{m}\left.\left[e\left(\frac{\mu\nu}{2m}\right)+e\left(-\frac{\mu\nu}{2m}\right) \right]g_\nu(\tau)
  \right), \nonumber \\
g_\mu\vert_{2\frac{1}{2}}\,T&=&e\left(\frac{\mu^2}{4m}\right)g_\mu(\tau). \nonumber
\end{eqnarray}
$A_\mathrm{s}$ and $A_\mathrm{p}$ can straightforwardly be
determined to be:
\begin{eqnarray}
\label{eq:As}
A_\mathrm{s}&=&\frac{m+1}{8}, \\
\label{eq:ap}
A_\mathrm{p}&=&-\frac{1}{2}S(m)-\sum_{\nu=0}^m \left(\left(\frac{\nu^2}{4m} \right)\right),
\end{eqnarray}
where $S(m)$ is equal to the number of times $\frac{\nu^2}{4m}\in
\mathbb{Z}$ for $\nu\in \left[0,m \right]$, which can be shown to be
equal to $\left\lfloor\frac{b+2}{2} \right\rfloor$, with $b$ the largest integer whose
square divides $m$ and $\lfloor \cdot \rfloor$ the floor function. The sum over $\nu$ in Eq. (\ref{eq:ap}) can be related to a sum over class numbers
$\cite{eichler}$, which is more convenient when one wants to evaluate
the sum for large $m$. 

We evaluate now $A_\mathrm{s}$. By Eq. (\ref{eq:gtransform}) we get
\begin{eqnarray}
&\chi_\mathbf{M}(S)&=\frac{1}{\sqrt{2mi}}\left(1-e\left(\frac{m}{2}\right)+\sum_{\mu=1}^m
e\left(\frac{\mu^2}{2m}\right)+e\left(-\frac{\mu^2}{2m}\right)\right), \\
&\chi_\mathbf{M}(ST)&=\frac{1}{\sqrt{2mi}}\left(
1-e\left(\frac{3m}{4}\right) + \sum_{\mu=1}^m
e\left(\frac{3\mu^2}{4m}\right)+e\left(-\frac{\mu^2}{4m}\right)
\right). \nonumber
\end{eqnarray}
These sums can be calculated using the analysis of quadratic Gauss
sums $G(n,m)=\sum_{r=1}^m e\left(\frac{nr^2}{m}\right)$
\cite{apostol:1976}. The relevant Gauss sums in this case are
\begin{eqnarray}
G(1,2m)&=&\left\{\begin{array}{rr}\sqrt{2m}(1+i), & m=0 \mod 2, \\ 0, &
m=1 \mod 2, \end{array}\right. \\
G(3,4m)&=&\left\{\begin{array}{rr} 2\sqrt{3m}(1+i), & m=0 \mod 3, \\ 2\sqrt{m}(1-i), & m=1 \mod 3, \\ -2\sqrt{m}(1-i), &
m=2 \mod 3. \end{array}\right. \nonumber
\end{eqnarray}
Then we obtain for $\chi_\mathbf{M}(S)$ and $\chi_\mathbf{M}(ST)$ 
\begin{eqnarray}
&\chi_\mathbf{M}(S)&=\left\{ \begin{array}{rr}
  e\left(-\frac{1}{8}\right), & m=0 \mod 2, \\ 0, & m=1 \mod
  2, \end{array} \right. \\
&\chi_\mathbf{M}(ST)&=\left\{ \begin{array}{rr}
  e\left(-\frac{1}{12}\right), & m=0 \mod 3, \\ e\left(-\frac{1}{4}\right), & m=1 \mod
  3, \\
  0, & m=2 \mod 3.
 \end{array} \right. \nonumber 
\end{eqnarray}
Inserting this in Eq. (\ref{eq:conttrace}) with $w=2\frac{1}{2}$, we find  
\be
A_\mathrm{e}=\left\{ \begin{array}{rr}
  -\frac{1}{4}, & m = 0\mod 2, \\ 0, & m=1 \mod
  2, \end{array} \right. + \left\{ \begin{array}{rr}
  -\frac{1}{3}, & m = 0 \quad \mod 3, \\ 0, & m=1,2 \mod
  3. \end{array} \right.
\ee
Thus, our final result for
$\mathrm{dim}\, S_{2\frac{1}{2},\mathbf{M}}(\Gamma(4m)^*)$ is:
\begin{eqnarray}
\label{eq:dimcusp}
&&\mathrm{dim}\, S_{2\frac{1}{2},\mathbf{M}}(\Gamma(4m)^*) =\frac{m+1}{8}-\frac{1}{2}S(m)-\sum_{\nu=0}^m
\left(\left(\frac{\nu^2}{4m} \right)\right) \\
&&+\left\{ \begin{array}{cc}
  -\frac{1}{4}, & m = 0\mod 2, \\ 0, & m=1 \mod
  2, \end{array} \right. + \left\{ \begin{array}{cl}
  -\frac{1}{3}, & m = 0 \mod 3, \\ 0, & m=1,2 \mod
  3. \end{array} \right. \nonumber 
\end{eqnarray}
The right-hand side grows linearly with $m$ for large $m$, it acquires
its first non-vanishing value for $m=5$. The quantities
$A_\mathrm{s}$, $A_\mathrm{e}$ and $A_\mathrm{p}$ behave differently
when $m$ is increased. Eq. (\ref{eq:As}) shows that $A_\mathrm{s}$
grows linearly with $m$ and $A_\mathrm{p}$ as $\sqrt{m}$. The absolute
value of $A_\mathrm{e}$ is always $<1$ and completes
$A_\mathrm{s}+A_\mathrm{p}$, such that the total sum is an integer. 

Ref. \cite{skoruppa:1984} evaluates $\chi_\mathbf{M}(\gamma)$ by
a decomposition of the representation $\mathbf{M}$ into
irreducible representations, and an evaluation of the characters of
the irreducible representations. The final dimension formulas are rather intricate
and involve several sums over integers and arithmetic
functions. The values obtained for
$\dim\,S_{2\frac{1}{2},\mathbf{M}}(\Gamma(4m)^*)$,
$m=1\dots 14$ by these formulas are identical to those obtained by
Eq. (\ref{eq:dimcusp}). The equivalence is however not proven for
general $m$.   


The dimension of the space of elliptic genera or weak Jacobi forms is
given by the number of polar coefficients $p(m)$ minus the number of
constraints (\ref{eq:j(m)}). The number $p(m)$ can be calculated as a function of $m$ \cite{eichler}
\be
p(m)=\sum_{\nu=0}^{m}\left\lceil\frac{\nu^2}{4m} \right\rceil.
\ee
The evaluation of $p(m)$ is less elaborate when the sum over $\nu$ is
rewritten using the functions $S(m)$ and $((x))$ introduced in Eq. (\ref{eq:ap}). This
gives for $p(m)$ \cite{eichler} 
\begin{eqnarray}
\label{eq:p(m)}
p(m)&=&\sum_{\nu=0}^m \left\{
\frac{\nu^2}{4m}+\frac{1}{2}-\left(\left(\frac{\nu^2}{4m}\right)\right)
\right\}-\frac{1}{2}S(m), \\
&=&\frac{m^2}{12}+\frac{5m}{8}+\frac{13}{24}-\frac{1}{2}S(m)-\sum_{\nu=0}^m
\left(\left(\frac{\nu^2}{4m} \right)\right). \nonumber
\end{eqnarray}
A comparison with (\ref{eq:dimcusp}) shows that the ratio of $p(m)$
and the number of constraints grows linearly with the central
charge. 

\begin{table}[h]
\caption{The number of polar coefficients $p(m)$ and the number of
  constraints on these coefficients, $\dim S_{2\frac{1}{2},\mathbf{M}}(\Gamma(4m)^*)$, for
  $\mathcal{N}=(2,2)$ elliptic genera as a function of $m$.}
\label{tab:N=2dimensions}
\begin{center}
\begin{tabular}{|l||r|r|r|r|r|r|r|r|r|r|r|r|r|r|}
\hline
$m$ & 1 & 2 & 3 & 4 & 5 & 6 & 7 & 8 & 9 & 10 & 11 & 12 & 13 & 14\\
\hline 
$p(m)$ & 1 & 2 & 3 & 4 & 6 & 8 & 9 & 11 & 13 & 16 & 18 & 21 & 23 & 27 \\
\hline
$\dim S_{2\frac{1}{2},\mathbf{M}}(\Gamma(4m)^*)$ & 0 & 0 & 0 & 0 & 1 &
1 & 1 & 1 & 1 & 2 & 2 & 2 & 2 & 3 \\
\hline
\end{tabular}
\end{center}
\end{table} 

The number of polar coefficients and the constraints on the
polar spectrum are listed in Table \ref{tab:N=2dimensions}. A
general expression for $\mathrm{dim}\,\tilde J_{0,m}$ is obtained by
inserting Eqs.  (\ref{eq:dimcusp}) and (\ref{eq:p(m)}) in (\ref{eq:j=p-dim}), 
\begin{eqnarray}
\label{eq:j(m)}
\mathrm{dim}\,\tilde J_{0,m}&=&\frac{m^2}{12}+\frac{m}{2} + \frac{5}{12}+ \left\{\begin{array}{ll}\frac{1}{4}, & m=0 \mod 2, \\
0, & m=1 \mod 2, \end{array} \right. \\
&& + \left\{\begin{array}{ll}\frac{1}{3}, & m=0 \mod 3, \\
0, & m=1,2 \mod 3. \end{array} \right. \nonumber
\end{eqnarray}
This result is identical to the dimension formula calculated in
Ref. \cite{eichler}. There, the dimension formula is derived by a study
of the Taylor expansion in $z$ of $\phi(\tau,z)$. This approach
determines that $\mathrm{dim}\,\tilde J_{0,m}$ is equal to
\be
\label{eq:dimj}
\mathrm{dim}\,\tilde J_{0,m}=\sum_{\nu=0}^m \mathrm{dim}\,M_{2\nu}(\Gamma),
\ee
where $M_{2\nu}(\Gamma)$ is the space of modular forms of weight
$2\nu$.  The dimension of $M_{2\nu}(\Gamma)$ is given by \cite{eichler}
\be
\label{eq:dimM}
\dim\,M_{2\nu}=\frac{2\nu+5}{12}-\frac{1}{3}\chi_3(2\nu-1)-\frac{1}{4}\chi_4(2\nu-1),
\ee
where $\chi_3$ and $\chi_4$ are the non-trivial Dirichlet characters
modulo three and four. Substitution of this expression in
Eq. (\ref{eq:dimj}) and evaluation of the sum over $\nu$ results
  precisely in Eq. (\ref{eq:j(m)}) \cite{eichler}.

The technique to calculate $\dim\,\tilde J_{0,m}$, described in the
previous sections, can be easily generalized to other spaces of Jacobi
forms. For example, one can determine the dimensions of spaces of Jacobi (cusp) forms
with weight $k\geq 3$, by requiring the vector $h_\mu(\tau)$ to be a
holomorphic (cusp) form with weight $k-\frac{1}{2}$. The described
techniques let us calculate the dimension of the appropriate
spaces. Dimensions of such Jacobi forms can also be obtained by a  
trace formula for Jacobi forms \cite{Skoruppa:1988, Skoruppa:1989,
  Skoruppa:2007} or  the Riemann-Roch theorem \cite{Kramer:1991}.      

Another generalization is the calculation of dimensions of spaces of
weak Jacobi forms with general weight, $\dim \tilde J_{k,m}$. This
might be relevant for physics, for example in the context of modular
differential equations for superconformal characters
\cite{Gaberdiel:2008pr}. Here we concentrate on $k$ even. Generically,
the polar coefficients might not determine the vector-valued form
$h_\mu(\tau)$ uniquely. When the weight is positive,
$k-\frac{1}{2}>0$, holomorphic forms might exist and some
non-polar terms must be specified as well to fix the
$h_\mu(\tau)$. The number of these coefficients is given by $\dim\,
M_{k-\frac{1}{2},\mathbf{\bar M}}(\Gamma(4m)^*)$. Of course,
a number of obstructions to finding $h_\mu(\tau)$ might still exist. This number
is given by $\dim\, S_{2\frac{1}{2}-k,\mathbf{M}}(\Gamma(4m)^*)$. As a
result, $\dim\,\tilde J_{k,m}$ is still given by the Selberg trace
formula 
\be
\label{eq:dimjkm}
\dim\,\tilde
J_{k,m}=p(m)-\left(\dim\, S_{2\frac{1}{2}-k,\mathbf{M}}(\Gamma(4m)^*)-\dim\,
M_{k-\frac{1}{2},\mathbf{\bar M}}(\Gamma(4m)^*)\right). 
\ee
Using the equations in Section \ref{sec:contcusp}, one can readily
evaluate the right hand side of (\ref{eq:dimjkm}). One finds
\begin{eqnarray}
\dim\,\tilde
J_{k,m}&=&\frac{m^2}{12}+\frac{5m}{8}+\frac{13}{24}-\frac{3-2k}{24}(m+1)
\nonumber \\
&&-\frac{1}{4}\left\{\begin{array}{cl} -1 & m=0 \mod 2,\,\, k=0 \mod
    4, \\ 1 & m=0 \mod 2,\,\, k=2 \mod 4,\\ 0 &
    \mathrm{else}, \end{array} \right. \\
&&-\frac{1}{3}\left\{\begin{array}{cl} -1 & m=0 \mod 3,\,\, k=0 \mod
    6, \\ 1 & m=0 \mod 3,\,\, k=2 \mod 6,\\ 1 & m=1 \mod 3,\,\, k=2
    \mod 6,\\ -1 & m=1 \mod 3,\,\, k=4 \mod 6,\\ 0 &
    \mathrm{else}. \end{array} \right. \nonumber
\end{eqnarray}
This is in agreement with the formulas given in \cite{eichler}. Note
that in the case of $k=2$, both $\dim\,
S_{2\frac{1}{2}-k,\mathbf{M}}(\Gamma(4m)^*)$ and $\dim\,
M_{k-\frac{1}{2},\mathbf{\bar M}}(\Gamma(4m)^*)$ might be
non-zero. The dimension formula for $\dim\,\tilde J_{2,m}$ does not
provide us the number of polar and non-polar coefficients which need
to be fixed.

\section{Constraints on $\mathcal{N}=(4,0)$ spectra}
\setcounter{equation}{0}
\label{sec:constraints04}

This section considers the constraints on the polar spectrum of $\mathcal{N}=(4,0)$
SCFT's by the modular symmetries of the $\CN=(4,0)$ elliptic
genus. The physical properties of this SCFT determine that
the elliptic genus is a mild generalization of a Jacobi form.
As in Section \ref{sec:egandjf}, a quasi-periodicity property can
be derived from a spectral flow symmetry, leading to a theta function
decomposition. The theta functions involved are however     
sums over a higher dimensional, non-definite, integral lattice
$\Lambda$, which leads to a vector $z^a$ of elliptic variables. The
representation of $\tilde \Gamma/\Gamma(4m)^*$ associated with the
vector-valued modular forms is also more intricate.  

The constraints are determined by the technique outlined in the
previous sections for weak Jacobi forms. The Taylor expansion, which is used in
\cite{eichler} to  determine the dimension formulas, is not 
useful in this context for several reasons. The main complications are
that the theta functions generically do not cancel the poles of the
vector-valued modular forms, the higher dimensional lattice $\Lambda$,
and that the $\CN=(4,0)$ elliptic genus depends on
$\tau$ as well as $\bar \tau$.  

As a final example of the calculation of constraints, the application to the partition
functions of $\mathcal{N}=4$ topologically twisted Yang-Mills on
four-manifolds is discussed \cite{Vafa:1994tf}. The case of
$\mathbb{CP}^2$ is worked out in some detail. The modular properties of the
vector-valued modular form, obtained from the $(4,0)$ elliptic genus,
closely resembles the properties of the set of gauge theory partition
functions for the groups $SU(N)/\mathbb{Z}_N$.  
 
The following paragraphs explain the appearance and relevance of these
generalizations of Jacobi forms in physics. Readers who are only
interested in the mathematical side of the discussion, might want to
go directly  to Eq. (\ref{eq:modtransformchi}).    
 
The main physical relevance of $\CN=(4,0)$ elliptic genera is the
fact that the degeneracies of M-theory black holes are enumerated by such
partition functions \cite{Maldacena:1997de, Minasian:1999qn,
  Gaiotto:2006wm, Kraus:2006nb, deBoer:2006vg}. Knowledge of the constraints on
the polar degeneracies is useful when one wants to determine explicit
expressions for such $\CN=(4,0)$ elliptic genera, as is done by
Refs. \cite{Gaiotto:2006wm, Gaiotto:2007cd}. Ref. \cite{Denef:2007vg}
derives an identical partition function from the viewpoint of IIA string theory. The $\CN=(4,0)$ SCFT
arises as the boundary conformal field theory in the near-horizon
geometry of an M-theory black hole. The near-horizon geometry is given
by AdS$_3\otimes S^2\otimes X$, where $X$ is a six-dimensional
Calabi-Yau. The black hole is sourced by an M5-brane with world-volume
fluxes, which wraps an ample divisor in $X$ and the boundary torus
$T^2$ of AdS$_3$. The Poincar\'e dual of
the divisor is $P=p^a \alpha_a$; the $\alpha_a$'s form a basis
of $H^2(X,\mathbb{Z})$ with $a=1\dots b_2$. The SCFT is obtained as a reduction of
the six-dimensional low-energy degrees of freedom to $T^2$. To this end, the typical length scale of the torus is
required to be much larger than that of $X$, and the M5-branes must be scarce 
in $X$ to avoid gravitational effects. Due to the amount
of supersymmetry, the trace over the Ramond sector needs an insertion
of $F^2$ to be non-vanishing \cite{Maldacena:1999bp}. The $\CN=(4,0)$ elliptic
genus $Z(\tau,\bar \tau,z)$ is defined as
\be
\label{eq:n4ellgenus}
Z(\tau, \bar \tau, z)=\mathrm{Tr}_\mathrm{R}\, \frac{1}{2}F^2
  (-1)^{F+p\cdot J_0} q^{L_0-\frac{c_L}{24}} \bar q^{\bar
  L_0-\frac{c_R}{24}} y^{J_0},
\ee
where $y^{J_0}=e(z\cdot J_0)$. The operators $J_{0,a}$ are the generators
of $b_2$ $U(1)$-charges. The charges are denoted by $q_a$ and represent the M2-brane charge of
the black hole. An electric charge vector $q$ is valued in the shifted dual lattice
$\Lambda^*+p/2$, where $\Lambda^*$ is the dual lattice of the magnetic
charge lattice $\Lambda$. The shift is a consequence of the
Freed-Witten anomaly \cite{Freed:1999vc}. An element $k \in \Lambda$ is an element
of $H^2(X,\mathbb{Z})$. The pull back of the inclusion map $i: P
\hookrightarrow X$ provides a non-degenerate integral inner-product $D$ on
$H^2(X,\mathbb{Z})$ 
\be
\label{eq:D}
D: H^2(X,\mathbb{Z})\otimes H^2(X,\mathbb{Z})\to \mathbb{C}, \qquad  
D(\rho,\sigma)=\int_P i^*\rho\wedge i^*\sigma=\int_X \rho \wedge \sigma \wedge P.
\ee
The quadratic form written as a matrix is $d_{ab}=d_{abc}p^c$, where
$d_{abc}=\int_X \alpha_a \wedge \alpha_b \wedge \alpha_c$ is the triple
intersection number of four-cycles. The inner-product on the dual
lattice is $d^{ab}=(d_{ab})^{-1}$. The dimension of this lattice is
the second Betti number $b_2$ and the signature is $(1,b_2-1)$. 

The central charges $c_L$ and $c_R$ of the SCFT are given by 
\be
c_L= p^3 + \frac{1}{2}c_2\cdot p, \qquad c_R= p^3 + c_2\cdot p,
\ee
where $c_2$ is the second Chern class of $X$. The zero-point energies
and the central charges determine that the momentum on the torus can
take the following values  
\be
\label{eq:momentum}
L_0-\bar L_{0}-\frac{c_L-c_R}{24}=\frac{p^3}{8}+\frac{c_R}{24} \mod \mathbb{Z}.
\ee
We denote the left-hand side of this equation by $-q_{\bar 0}$, where
$q_{\bar 0}$ is the anti-D0-brane charge in IIA string theory. The
quantity $q_{\bar 0}+\frac{1}{2}q^2$ is positive for black holes. The charges $p^a$ and $q_a$ correspond, respectively, to the D4-brane and
D2-brane charges in IIA string theory. The insertion of $F^2$ projects
the trace on $\frac{1}{2}$-BPS states. For these states
$L_0-\frac{c_L}{24}$ satisfies  
\begin{equation}
\label{eq:L0halfBPS}
L_0-\frac{c_L}{24}=\frac{1}{2}q_+^2,
\end{equation} 
where the subscript $+$ denotes projection to the positive definite
sublattice. This condition implies a ``heat equation''
\be
\left(\partial_{\tau}+{\textstyle\frac{i}{4\pi}}\partial^2_{z_+}\right)Z(\tau,\bar \tau, z)=0.
\ee

The weight of $Z(\tau,\bar \tau, z)$ is $(\frac{1}{2},-\frac{3}{2})$,
which is a consequence of the space-time momenta and the insertion of
$F^2$. Since the quantity in Eq. (\ref{eq:momentum}) is in general not
an integer, $\Gamma$-transformations are accompanied by phase factors
\begin{eqnarray}
\label{eq:modtransformchi}
&Z\vert_{\left(\frac{1}{2},-\frac{3}{2}\right)} S =& \varepsilon(S)\, e\left(\frac{z_+^2}{2\tau}+\frac{z_-^2}{2\bar
\tau}\right)Z(\tau, \bar \tau,z), \\
&Z\vert_{\left(\frac{1}{2},-\frac{3}{2}\right)} T =& \varepsilon(T)\,
Z(\tau, \bar \tau,z).
\end{eqnarray}
Using (\ref{eq:momentum}) and that $\frac{1}{6}p^3+\frac{1}{12}c_2\cdot p \in \mathbb{Z}$ by the index
formula, we deduce that $\varepsilon(T)=e\left(-\frac{c_2\cdot
  p}{24}\right)$. Consistency of modular transformations requires that
the $S$-transformation is also accompanied by a unitary pre-factor
$\varepsilon(S)$. Since both $(ST)^3=-\mathbf{1}$ and $S^2=-\mathbf{1}$ leave
$\tau$ invariant, $\varepsilon(S)=\varepsilon(T)^{-3}$. The phases for
general $\gamma$ are denoted by $\varepsilon(\gamma)$.

As in Section \ref{sec:egandjf}, this SCFT also contains a spectral
flow symmetry which implies a quasi-periodicity of the partition
function as a function of $z^a$. This symmetry relates a state with charge $q$ to a state
with charge $q + k$, where $k\in
\Lambda$. Spectral flow determines in this way equivalence classes for the charges $q\in
p/2+\Lambda^*$. The collection of all coset representatives forms the
discriminant group $\Lambda^*/\Lambda$, which is finite and abelian. Its order 
$|\Lambda^*/\Lambda|$ divides $D=\det d_{ab}$. The representatives
$\mu$  are glue vectors and chosen such that they have minimal absolute
length $|\mu^2|$. Spectral flow as a symmetry of the spectrum, determines that
the degeneracies $c(q_{\bar 0}, q_a)$ depend only on the
equivalence class of $q_a$ in $\Lambda^*/\Lambda+p/2$ and
$q_{\bar   0}+\frac{1}{2}q^2$. The part of the spectrum
with $q_{\bar   0}+\frac{1}{2}q^2<0$, is the polar spectrum in this
case. The spectral flow or quasi-periodicity combined with (\ref{eq:L0halfBPS}), allows
us to perform a decomposition of $Z(\tau, \bar \tau, z)$ into a
vector-valued modular form $h_\mu(\tau)$ and theta-functions
$\Theta_{\mu}(\tau,\bar \tau, z)$  
\be
\label{eq:fourierdecompchi}
Z(\tau, \bar \tau, z)=\sum_{\mu \in \Lambda^*/\Lambda}
\overline{h_{\mu}( \tau)} \Theta_\mu(\tau, \bar \tau, z).
\ee
Note that this decomposition claims that $h_\mu(\tau)$ is a (weakly holomorphic)
function of $\tau$. All the dependence on $\tau$ and $z$ of $Z(\tau,
\bar \tau, z)$ is captured by the $\Theta_\mu(\tau,\bar \tau, z)$. The theta function
corresponds to the $U(1)^{b_2}$-sector of the theory, which are the so-called
singleton  degrees of freedom \cite{Witten:1998wy, Maldacena:2001ss,
  Moore:2004jv} from the point of view of AdS$_3$-supergravity. The expansions of
$h_{\mu}(\tau)$ and $\Theta_\mu(\tau, \bar \tau, z)$ are
\be
\label{eq:hmu}
h_{\mu}(\tau)=\sum_{n=0}^\infty c_{\mu}(n) q^{n-\Delta_\mu},
\ee
\begin{eqnarray}
\Theta_\mu(\tau, \bar \tau, z)=  \sum_{k \in \Lambda + p/2} (-1)^{p\cdot (k+\mu)} e\left(\tau(k+\mu)^2_+/2 + \bar\tau(k+\mu)^2_-/2 +(k+\mu)\cdot z \right),
\end{eqnarray}
where $\Delta_\mu\in \mathbb{Q}$ and generically positive. It is given by
\be
\label{eq:Delta}
\Delta_\mu=\frac{c_R}{24}-\left(\frac{\mu^2+p\cdot \mu}{2}-\left\lfloor
\frac{\mu^2+p\cdot \mu}{2} \right\rfloor \right),
\ee
such that $n-\Delta_\mu$ takes values in 
$\frac{c_2\cdot p}{24}+\frac{1}{2}(\mu+p/2)^2 \mod \mathbb{Z}$. 

The theta functions transform among each other under modular
transformations. If we define 
\begin{eqnarray}
\Theta_\mu\vert_{(b_2^+,b_2^-)}\,
\gamma&=&j(\gamma,\tau)^{-b^+_2/2}j(\gamma,\bar \tau)^{-b^-_2/2}e\left(-\frac{cz_+^2}{c\tau+d}-\frac{cz_-^2}{c\bar
  \tau+d}\right)  \nonumber \\
&&\times\Theta_\mu\left(\gamma(\tau),\gamma(\bar \tau),
\frac{z_+}{c\tau+d}+\frac{z_-}{c\bar \tau+d}\right),
\end{eqnarray}
then the $S$- and $T$-transformation are given by
\begin{eqnarray}
\label{eq:thetatransformII}
&\Theta_\mu \vert_{(b_2^+,b_2^-)}\,
S & =\frac{1}{\sqrt{|\Lambda^*/\Lambda|}}(-i)^{(b^+_2-b^-_2)/2}
e\left(-\frac{p^2}{4}\right) \sum_{\delta \in \Lambda^*/\Lambda} 
e(-\delta\cdot \mu)\Theta_{\delta}(\tau,\bar \tau, z), \nonumber \\
&\Theta_{\mu}\vert_{(b_2^+,b_2^-)}\, T& = e\left(\frac{(\mu+p/2)^2}{2} \right) \Theta_{\mu}(\tau ,\bar \tau , z),
\end{eqnarray}
with $b^+_2=1$ and $b^-_2=b_2-1$, because the signature of the lattice is $(1,b_2-1)$.
Note that $p^2$ is actually $p^3$, since the inner product is
given by $d_{abc}p^c$. However, since we are working here with
quadratic forms, we use the notation $p^2$. The transformations of
$\Theta_\mu$ for generic elements $\gamma \in \Gamma$ can be derived using the special
properties of these theta functions, following for example
\cite{ogg:1969}. \footnote{I am grateful to the referee for pointing
  out this reference to me.} Since $\mu^2\in \mathbb{Q}$ for every $\mu\in
\Lambda^*/\Lambda$, an integer $m$ exists such that
$\Theta_{\mu}\vert_{(b_2^+,b_2^-)}\, T^{4m} = \Theta_{\mu}(\tau ,\bar
\tau , z)$ for every $\mu\in
\Lambda^*/\Lambda$. If $\gamma=\left( \begin{array}{cc} a & b \\ c & d
\end{array}\right)$ with $a,d$ odd, and $c=0 \mod 4m$, then
\be
\Theta_\mu\vert_{(b_2^+,b_2^-)}\, \gamma = \varphi(\mu,\gamma)\,\Theta_{a\mu}(\tau,\bar
\tau, z),
\ee
with 
\be
\varphi(\mu,\gamma)=d^{-\frac{b_2}{2}}e\left({\textstyle\frac{d-1}{4}p^2+\frac{ab}{2}(\mu^2+\mu\cdot
  p)}\right)\sum_{\ell \in \Lambda/d \Lambda}e\left({\textstyle\frac{b}{2d}}\left(p/2+ \ell \right)^2 \right).
\ee  
The sum over $\ell \in \Lambda/d \Lambda$ can be evaluated using
Gauss sums, from which one can derive that $\Theta_\mu \vert_{(b_2^+,b_2^-)}\,
\gamma=\Theta_\mu (\tau,\bar \tau, z)$ with $\gamma \in \Gamma(4m)^*$ for
some $m$. 

%
%
%
%

The weight of the theta functions $\frac{1}{2}(1,b_2-1)$ works out
nicely with the weakly holomorphicity of $h_\mu(\tau)$. They have
weight $(-\frac{b_2}{2}-1,0)$ and transform under $S$ 
and $T$ as 
\begin{eqnarray}
h_\mu\vert_{-1-\frac{b_2}{2}}\, S&=&-\frac{1}{\sqrt{|\Lambda^*/\Lambda|}}
(-i)^{-b_2/2-1}\varepsilon(S)^* e\left(-\frac{p^2}{4}\right) \sum_{\delta \in
  \Lambda^*/\Lambda} e(-\delta\cdot 
\mu)h_\delta(\tau), \nonumber \\
\label{eq:transvectform}
h_\mu\vert_{-1-\frac{b_2}{2}}\, T&=&\varepsilon(T)^* e\left(\frac{(\mu+p/2)^2}{2}\right)
h_\mu(\tau). 
\end{eqnarray}
The additional $-$-sign, appearing in the $S$-transformation, is a
consequence of the unitary factor in (\ref{eq:thetatransformII}). 
The vector $h_\mu(\tau)$ has length $|\Lambda^*/\Lambda|$. Similarly
to the discussion around (\ref{eq:z-z}), some elements of the
vector are required to be identical. This is again deduced from the
transformation of $Z(\tau, \bar \tau, z)$ under 
$S^2=-\mathbf{1}$
\be
Z(\tau, \bar \tau, -z)=\varepsilon(S)^2 Z(\tau, \bar \tau,
z).
\ee
Moreover, $-\mathbf{1}$ acts on $\Theta_\mu(\tau,\bar \tau, z)$ by
\be
\Theta_\mu(\tau,\bar \tau,-z)=(-1)^{p^3}\Theta_\nu(\tau,\bar \tau, z), \qquad \nu=-\mu \mod \Lambda.
\ee
Since $-\mathbf{1}$ acts trivially on $h_\mu(\tau)$, these equations
determine that 
\be
\label{eq:h-mu}
h_{\mu}(\tau)=h_\nu(\tau), \qquad \nu=-\mu \mod \Lambda.
\ee
Thus the dimension $d$ of the vector $h_\mu(\tau)$ equals the number
of orbits in $\Lambda^*$ by the group $\Lambda\otimes
\mathbb{Z}_2$. Here $\Lambda$ acts additively and $\mathbb{Z}_2$
is multiplication by $\pm \mathbf{1}$. These orbits do not
naturally form a group. For a one-dimensional lattice with inner
product $\ell$, $d=\frac{1}{2}\ell+1$ for $\ell$ even, and
$d=\frac{1}{2}(\ell+1)$ if $\ell$ is odd. 

%
%

Since the individual $\Theta_\mu(\tau,\bar \tau, z)$ are forms of a congruence subgroup, and
$\varepsilon(\gamma)$ is just the multiplier system of a power of $\eta(\tau)$, the functions $h_\mu(\tau)$
are (weakly holomorphic) modular forms of a congruence subgroup $\Gamma(4m)^*$ for some
$m$. \footnote{Note that the level arising for the $\Theta_\mu(\tau,\bar \tau, z)$  might differ from the level of
this congruence subgroup, since the transformation properties are
slightly different.} This shows that the formulas (\ref{eq:dims})
and (\ref{eq:conttrace}), which we used to calculate the number of
constraints in Section \ref{sec:contcusp}, are also applicable
here. The $\Delta_\mu$ in the expansion (\ref{eq:hmu}) play the role
of $\lambda_j$ in (\ref{eq:conttrace}). The form of $\Delta_\mu$ in
(\ref{eq:Delta}) indicates that $h_\mu(\tau)$ is naturally written as
$f_\mu(\tau)/\eta(\tau)^{c_R}$. The dimension of the space of 
weakly holomorphic $h_\mu(\tau)$ can then be calculated as the space of
holomorphic $f_\mu(\tau)$ by (\ref{eq:dims}), if $c_R \geq b_2+2$. These $f_\mu(\tau)$
might capture interesting data as in the case of $SU(2)$ $\CN=4$
Yang-Mills theory on $\mathbb{CP}^2$, see the last part of this
section for more details. 

It is interesting to compare the ratio between the number of polar
coefficients and constraints. The number of polar terms
$p(\mathbf{M},c_R)$ is given by 
\be
\label{eq:polar}
p(\mathbf{M},c_R)=\sum_{\mu\in \Lambda^*/\Lambda \otimes \mathbb{Z}_2} \lceil
\Delta_\mu\rceil=-\frac{1}{2}S(\mathbf{M})+\sum_{\mu \in\Lambda^*/\Lambda \otimes \mathbb{Z}_2}\left\{\Delta_\mu-((\Delta_\mu))+\frac{1}{2}
\right\}, 
\ee
where $S(\mathbf{M})$ is the number of terms $\Delta_\mu$ which take
their value in $\mathbb{Z}$. Note that $p(\mathbf{M},c_R)$ grows as
$\frac{c_R}{24}d$ for large $c_R$, whereas the obstruction space grows as
$\frac{4+b_2}{24}d$. Therefore, as in the case of $\mathcal{N}=(2,2)$
elliptic genera, the ratio between the number of polar terms and
the obstructions is proportional to the central charge. Another
resemblance is the appearance of the quantity $A_\mathrm{p}$ in the
number of polar terms. The contribution of the triple intersection
number to $c_R$ leads to many polar coefficients, therefore the
constraints are not very restrictive. A general analysis can be carried out, 
but instead only some comments on specific examples are given here. For one
M5-brane on the hyperplane section of the quintic Calabi-Yau (considered by
\cite{Gaiotto:2006wm, Gaiotto:2007cd}) one finds seven polar terms and no
constraints; for two M5-branes the number of polar terms has already
increased to 36 and then also one constraint is
present. Ref. \cite{Gaiotto:2007cd} analyzes a number of other
situations where an M5-brane wraps a surface in a Calabi-Yau with
$b_2=1$. Interestingly, for an
M5-brane wrapping the hyperplane section of the bicubic in
$\mathbb{CP}^5$ is reported that six basis elements suffice to
determine the elliptic genus, whereas the number of polar coefficients
is seven. Indeed, one can show with the above described technique, that
one constraint is present for this example. In agreement with
\cite{Gaiotto:2007cd}, no constraints are found in the other   
examples worked out there.

The constraints are stronger in the case of twisted $\mathcal{N}=4$
supersymmetric Yang-Mills theory on a four-manifold $M$. The coupling
constant $g$ and the $\theta$-angle of the theory are conveniently combined in the
complex coupling constant $\tau=\frac{\theta}{2\pi}+\frac{4\pi
  i}{g^2}$. This theory is invariant under a strong-weak coupling
duality ($S$-duality) \cite{Montonen:1977sn}, except that
the gauge group $G$ of the theory is exchanged with the dual magnetic group
$\hat G$ \cite{Goddard:1976qe}. For example, the dual group of $SU(N)$
is $SU(N)/\mathbb{Z}_N$. When certain conditions are
satisfied, the partition function is the generating function for the
Euler numbers of instanton moduli spaces \cite{Vafa:1994tf}.  The $S$-duality
manifests itself as modular behavior of the partition function of
the twisted theory.  Ref. \cite{Vafa:1994tf}
explains that the partition functions for gauge group $SU(N)/\mathbb{Z}_N$, with
different 't Hooft fluxes valued in $\mathbb{Z}/\mathbb{Z}_N$,
transform among each other as a vector-valued modular form with
weight $-\chi(M)/2$. The partition function of $SU(N)$ is given by the one of
$SU(N)/\mathbb{Z}_N$ with trivial 't Hooft flux, multiplied by
$N^{-1+b^1}$ (with $b^1$ the first Betti number). 

Remarkably, the transformation properties
(\ref{eq:transvectform}) closely resemble the transformation
properties of the $SU(N)/\mathbb{Z}_N$ partition functions of twisted $\mathcal{N}=4$
Yang-Mills theory. Both the $S$- and $T$-transformation are compatible.
This resemblance shows that the $SU(N)/\mathbb{Z}_N$ partition
functions can be combined into a single partition function by adding
$U(1)$ degrees of freedom. This gives the partition function of the
theory with gauge group $U(N)$, whose magnetic group is $U(N)$ as well. 
The space of constraints is more restrictive in this
situation than for $\CN=(4,0)$ SCFT's, since the number of polar
degeneracies is much smaller now the contribution to the central
charge of the triple intersection number vanishes. In the following,
the number of polar terms and obstructions are calculated for the case
of $\mathbb{CP}^2$, which gives the dimension of the space of weakly
holomorphic modular forms with the required transformation properties. A
determination of the polar coefficients, which might involve the so-called gap
condition \cite{Vafa:1994tf}, is not attempted.


Since $b_2=1$ for $\mathbb{CP}^2$, the 
lattice $\Lambda$ is one-dimensional and the theta function is
holomorphic in $\tau$. Consequently, $h_\mu(\tau)$ has weight $-\frac{3}{2}$. The
second Chern class of $\mathbb{CP}^2$ is $3J^2$, with $J$ the hyperplane 
class. The central charge $c_R$ from the SCFT reduces to the combination $\chi(\mathbb{CP}^2) N=3N$. The unitary factor $\varepsilon(T)$ is then given by 
$\varepsilon(T)=e(\frac{N}{8}+\frac{c_R}{24})=e(\frac{N}{4})$. Note
that the index formula cannot be used in this situation, and
that therefore $\varepsilon(T)\neq e(-\frac{c_2\cdot N}{24})$. The
theta functions $\Theta_{N,\mu}$ are given by
\be
\Theta_{N,\mu}(\tau,z)=\sum_{k\in \mathbb{Z}}
e\left(\frac{\tau}{2N}\left(\frac{N}{2}+\mu+kN \right)^2 +
\left(\frac{N}{2}+\mu+kN \right)(z+\textstyle{\frac{1}{2}})\right). 
\ee
From the transformation properties of $\Theta_{N,\mu}$ follows that
$h_\mu(\tau)$ transforms as
\begin{eqnarray}
h_\mu\vert_{-1-\frac{b_2}{2}}\, S &=&-\frac{1}{\sqrt{N}}
(-i\tau)^{-\frac{3}{2}} e\left(\frac{N}{2}\right) \sum_{\nu \mod
N} e\left(-\frac{\mu\nu}{N}\right)h_\nu(\tau), \nonumber \\
\label{eq:transvectformCP}
h_\mu\vert_{-1-\frac{b_2}{2}}\, T &=& e\left(-\frac{N}{4}+\frac{1}{2N}\left(\mu+\frac{N}{2}\right)^2\right)
h_\mu(\tau). 
\end{eqnarray}
The functions satisfy moreover
$\Theta_{N,\mu}(\tau,-z)=(-)^N\Theta_{N,-\mu}(\tau,z)$ and
$h_\mu(\tau)=h_{-\mu}(\tau)$, such that $h_\mu(\tau)$ can be reduced to
a vector of length $\frac{N}{2}+1$ if $N$ is even and $\frac{N+1}{2}$
if $N$ is odd. The elements $h_\mu(\tau)$ are forms of $\Gamma(2N)$ for $N$
even and otherwise $\Gamma(8N)$. The number of polar terms $p(N)$
is given by 
\be
p(N)=\sum_\mu \left\lceil\frac{N}{8}-\left(\frac{\mu^2}{2N}+\frac{\mu}{2}-\left\lfloor
    \frac{\mu^2}{2N}+\frac{\mu}{2} \right\rfloor\right) \right\rceil.
\ee
One can straightforwardly determine the properties of the obstruction
forms $g_\mu(\tau)$. They have weight $3\frac{1}{2}$, we denote their
representation again by $\mathbf{M}$. A closed expression can be given
for the dimension of the space of forms which satisfy the required properties
for $h_\mu(\tau)$. Table \ref{tab:cp2dimensions} presents the
number of polar coefficients $p(N)$
and constraints on the polar spectrum for $N=1\dots 10$. For large $N$,
$p(N)$ grows as $\frac{1}{16}N^2$ and the number of constraints as $\frac{5}{48}N$.
\begin{table}[h]
\caption{For $U(N)$ gauge theory on $\mathbb{CP}^2$, the number of polar coefficients $p(N)$ and constraints on the polar
  spectrum $\dim S_{3\frac{1}{2},\mathbf{M}}$ are listed for $N=1\dots 10$.}
\label{tab:cp2dimensions}
\begin{center}
\begin{tabular}{|l||r|r|r|r|r|r|r|r|r|r|}
\hline
$N$ & 1 & 2 & 3 & 4 & 5 & 6 & 7 & 8 & 9 & 10\\
\hline 
$p(N)$ & 1 & 1 & 1 & 1 & 3 & 4 & 4 & 5 & 7 & 8 \\
\hline
$\dim S_{3\frac{1}{2},\mathbf{M}}$ & 0 & 1 & 0 & 0 & 1 & 1 & 0 & 0 & 1 &
2\\
\hline
\end{tabular}
\end{center}
\end{table} 

A first observation is that $\dim S_{3\frac{1}{2},\mathbf{M}}$ does grow
more irregularly than the dimension of the space of cusp forms presented
in Table \ref{tab:N=2dimensions}. Table \ref{tab:cp2dimensions} confirms
earlier results for  $N=1$ and $N=2$, which are derived using the Weil conjectures
\cite{gottsche:1990, Yoshioka:1994}. For $N=1$, the
space of potential partition functions is one-dimensional, therefore, $h_0(\tau)$ must be proportional to
$\eta(\tau)^{-3}$. This agrees with the computation in
Ref. \cite{gottsche:1990}. For $N=2$, the table teaches us that no
weakly holomorphic function $h_\mu(\tau)$ exists with the required
properties. The obstruction form is given explicitly in
\cite{Manschot:2007ha}. The fact that no modular weakly holomorphic partition
function exists, is consistent with the known partition function
calculated in  \cite{Yoshioka:1994}. This non-modular partition
function has the form $f_\mu(\tau)/\eta(\tau)^6$ \cite{Vafa:1994tf}, where
$f_\mu(\tau)$ are regularized Eisenstein series of weight
$\frac{3}{2}$. The Fourier coefficients $c_\mu(n)$ of $f_\mu(\tau)$ are
the number of equivalence classes of all positive definite forms, with
discriminant $-4n$ and $-4n+1$ for $\mu=0$ and $\mu=1$ respectively. A connection between class numbers and rank two bundles over
$\mathbb{CP}^2$ was earlier established by \cite{Klyachko:1991}. The
holomorphic $f_\mu(\tau)$ are not modular covariant, but can be made
so by a non-holomorphic addition \cite{zagier:1975, hirzebruch:1976},
similar to the way $E_2(\tau)$ can be made modular covariant. It is
unclear how and if this holomorphic anomaly extends to larger
values of $N$. Ref. \cite{Minahan:1998vr} explains how it does if the
four-manifold is $\frac{1}{2}$K3.  

The  table shows that appropriate weakly holomorphic forms
exists and are unique up to an overall factor for $N=3$ and $N=4$. One
can show that for $N=3$, the $h_\mu(\tau)$ are given by 
\be
\label{eq:hcp2}
h_\mu(\tau)=\frac{1}{2}\frac{\theta^5_2(\tau)\Theta_{3,\mu}\begin{bmatrix}
      \textstyle{\frac{1}{2}} \\ 0 \end{bmatrix}(\tau)+\theta^5_3(\tau)\Theta_{3,\mu}\begin{bmatrix}
      0 \\ 0 \end{bmatrix}(\tau)+\theta^5_4(\tau)\Theta_{3,\mu}\begin{bmatrix}
      0 \\ \textstyle{\frac{1}{2}} \end{bmatrix}(\tau)}{\eta(\tau)^9},
\ee
where $\Theta_{N,\mu}\begin{bmatrix} a \\ b \end{bmatrix}(\tau)$ is defined by
\be
\Theta_{N,\mu}\begin{bmatrix} a \\ b \end{bmatrix}(\tau)=\sum_{k\in \mathbb{Z}}
e\left(\frac{\tau}{2N}\left(aN+\mu+kN \right)^2 + \left(aN+\mu+kN \right)b\right).
\ee
The overall factor of $\frac{1}{2}$ in (\ref{eq:hcp2}) is such that the first coefficient
of the expansion is one. For $N=4$, $h_\mu(\tau)$ can be written in the form
$f_\mu(\tau)/\eta(\tau)^{12}$, where $f_\mu(\tau)$ is a holomorphic
vector-valued modular form. 

It would be interesting to learn whether these functions give
the Euler numbers of the instanton moduli spaces on
$\mathbb{CP}^2$. This could shed light on the question of the holomorphic anomaly as well as the gap condition.
It would also be instructive to find out whether the Euler numbers for
$N=3$ and $N=4$ capture any special information, as is the case for
$N=2$ where they are related to class numbers. 

\section{Conclusion}
The dimension of the space of $\CN=(2,2)$ and $(4,0)$ elliptic genera
is shown to be equal to the number of independent degeneracies of the
polar spectra, minus a number of constraints. This excludes
certain polar spectra of SCFT's, since not all spectra are
consistent with modular transformations. The constraints are also of
interest from the point of view of AdS$_3$-gravity, since the
polar spectrum corresponds to the states which lie classically
below the cosmic censorship bound. A technique is presented to
calculate the number of constraints, which is generally applicable
in situations where weakly holomorphic modular forms appear. As an
additional example, $\CN=4$ Yang-Mills on $\mathbb{CP}^2$ is
discussed. This showed that weakly holomorphic modular forms do exist, which
satisfy the modular properties expected for $SU(3)$ and $SU(4)$
partition functions. In a similar way, the technique might
prove useful in future situations where one wants to determine 
specific modular forms.  

From the mathematical point of view, the space of weak Jacobi forms of
weight zero and index $m$ ($\tilde J_{0,m}$) is studied. A Jacobi form
can be decomposed in a vector-valued modular form $h_\mu(\tau)$ and
theta-functions $\theta_{m,\mu}(\tau,z)$ by quasi-periodicity. The
dimension of $\tilde J_{0,m}$ is calculated as the number of polar
coefficients of $h_\mu(\tau)$ minus the number of constraints imposed
by the existence of a space of cusp forms.  Also the generalization to
$\dim\,\tilde J_{k,m}$ is given. In addition, the used techniques are
applied to some generalizations of Jacobi forms, which arise in
physics.

\bigskip
\begin{center} {\bf Acknowledgments}\end{center}
I am thankful to M. R. Gaberdiel, T. Gannon, G. B. M. van der Geer and
G. W. Moore for comments and helpful discussions. I also would like to
thank the workshop on ``Three-Dimensional Quantum Gravity'' organized
by the McGill Center for High Energy Physics, the Department of
Physics and Astronomy of Rutgers University and the Institute for
Theoretical Physics of the ETH Z\"urich for their kind hospitality. My
research is financially supported by the Foundation of Fundamental
Research on Matter (FOM).

\providecommand{\href}[2]{#2}\begingroup\raggedright\endgroup


\begin{thebibliography}{1}
\bibitem{Cardy:1986ie}
  J.~L.~Cardy,
  {\it Operator Content Of Two-Dimensional Conformally Invariant Theories},
  Nucl.\ Phys.\  B {\bf 270} (1986) 186.

\bibitem{Aharony:1993zu}
  O.~Aharony, S.~Yankielowicz and A.~N.~Schellekens,
  {\it Charge Sum Rules In N=2 Theories},
  Nucl.\ Phys.\  B {\bf 418} (1994) 157
  [arXiv:hep-th/9311128].

\bibitem{Witten:1986bf}
  E.~Witten,
  {\it Elliptic  genera and quantum field theory},
  Commun.\ Math.\ Phys.\  {\bf 109} (1987) 525.

\bibitem{Eguchi:1988vra}
  T.~Eguchi, H.~Ooguri, A.~Taormina and S.~K.~Yang,
  {\it Superconformal Algebras and String Compactification on Manifolds with SU(N)
  Holonomy},
  Nucl.\ Phys.\  B {\bf 315} (1989) 193.

\bibitem{gritsenko-991}
V.~Gritsenko, {\it Elliptic genus of calabi-yau manifolds and jacobi
and siegel
  modular forms},  \href{http://xxx.lanl.gov/abs/math/9906190}{{\tt
  math/9906190}}.

\bibitem{Strominger:1996sh}
  A.~Strominger and C.~Vafa,
  {\it Microscopic Origin of the Bekenstein-Hawking Entropy},
  Phys.\ Lett.\  B {\bf 379} (1996) 99
  [arXiv:hep-th/9601029].

\bibitem{Maldacena:1997de}
  J.~M.~Maldacena, A.~Strominger and E.~Witten,
  {\it Black hole entropy in M-theory},
  JHEP {\bf 9712} (1997) 002
  [arXiv:hep-th/9711053].

\bibitem{Maldacena:1997re}
  J.~M.~Maldacena,
  {\it The large N limit of superconformal field theories and supergravity},
  Adv.\ Theor.\ Math.\ Phys.\  {\bf 2} (1998) 231
  [Int.\ J.\ Theor.\ Phys.\  {\bf 38} (1999) 1113]
  [arXiv:hep-th/9711200].

\bibitem{Cvetic:1998xh}
  M.~Cvetic and F.~Larsen,
  {\it Near horizon geometry of rotating black holes in five
  dimensions },
  Nucl.\ Phys.\  B {\bf 531} (1998) 239
  [arXiv:hep-th/9805097].

\bibitem{Dijkgraaf:2000fq}
  R.~Dijkgraaf, J.~M.~Maldacena, G.~W.~Moore and E.~P.~Verlinde,
  {\it A black hole farey tail},
  [arXiv:hep-th/0005003].

\bibitem{Manschot:2007ha}
  J.~Manschot and G.~W.~Moore,
  {\it A Modern Farey Tail,}
  arXiv:0712.0573 [hep-th].

\bibitem{niebur:1974}
D.~Niebur, {\it Construction of automorphic forms and integrals},
  Transactions of the Amer. Math. Soc. {\bf 191} (1974) 373--385.

\bibitem{gaberdiel:2008}
M.~R.~Gaberdiel, S.~Gukov, C.~A.~Keller, G.~W.~Moore, H.~Ooguri, {\it
  Extremal $\mathcal{N}=(2,2)$ 2D Conformal Field Theories and
  Constraints of Modularity,}   arXiv:0805.4216 [hep-th]. 

\bibitem{eichler}
M.~Eichler and D.~Zagier, {\em The Theory of Jacobi Forms}.
\newblock Birkh\"auser, 1985.

\bibitem{skoruppa:1984}
N.~P.~ Skoruppa, {\em \"Uber den Zusammenhang zwischen Jacobi-Formen und
  Modulformen halbganzen Gewichts}. Dissertation, Universit\"at Bonn, 1984.

\bibitem{Minasian:1999qn}
  R.~Minasian, G.~W.~Moore and D.~Tsimpis,
  {\it Calabi-Yau black holes and (0,4) sigma models},
  Commun.\ Math.\ Phys.\  {\bf 209} (2000) 325
  [arXiv:hep-th/9904217].

\bibitem{Gaiotto:2006wm}
  D.~Gaiotto, A.~Strominger and X.~Yin,
  {\it The M5-brane elliptic genus: Modularity and BPS states},
  JHEP {\bf 0708}, 070 (2007)
  [arXiv:hep-th/0607010].

\bibitem{Kraus:2006nb}
  P.~Kraus and F.~Larsen,
  {\it Partition functions and elliptic genera from supergravity},
  JHEP {\bf 0701} (2007) 002
  [arXiv:hep-th/0607138].

\bibitem{deBoer:2006vg}
  J.~de Boer, M.~C.~N.~Cheng, R.~Dijkgraaf, J.~Manschot and E.~Verlinde,
  {\it A farey tail for attractor black holes},
  JHEP {\bf 0611}, 024 (2006)
  [arXiv:hep-th/0608059].

\bibitem{Kawai:1993jk}
  T.~Kawai, Y.~Yamada and S.~K.~Yang,
  {\it Elliptic Genera And N=2 Superconformal Field Theory},
  Nucl.\ Phys.\  B {\bf 414} (1994) 191
  [arXiv:hep-th/9306096].

\bibitem{koblitz}
N.~Koblitz, {\em Introduction to Elliptic Curves and Modular Forms}.
\newblock Springer-Verlag, 1993.

\bibitem{knopp:1974}
I.~Knopp, Marvin, {\it Some new results on the Eichler cohomology of
  automorphic forms},   Bulletin of the Amer.
  Math. Soc. {\bf 80} (1974) 607-632.

\bibitem{Witten:2007kt}
  E.~Witten,
  {\it Three-Dimensional Gravity Revisited},
  arXiv:0706.3359 [hep-th].

\bibitem{sarnak:1990}
P.~Sarnak, {\em Some applications of modular forms}.
\newblock Cambridge University Press, 1990.

\bibitem{griffiths}
P.~Griffiths and J.~Harris, {\em Principles of Algebraic Geometry}.
\newblock John Wiley and Sons, 1978.

\bibitem{borcherds:1999}
R.~E.~Borcherds, {\it The Gross-Kohnen-Zagier theorem in higher dimensions},
  Duke Math. J. {\bf 97} (1999) 219--233.

\bibitem{Eholzer:1994ta}
  W.~Eholzer and N.~P.~Skoruppa,
  {\it Modular invariance and uniqueness of conformal characters},
  Commun.\ Math.\ Phys.\  {\bf 174} (1995) 117
  [arXiv:hep-th/9407074].

\bibitem{Bantay:2007}
  P.~Bantay and T.~Gannon,
  {\it Vector-valued modular functions for the modular group and the
    hypergeometric equation},
    arXiv:0705.2467 [math.NT].

\bibitem{zagier:1976}
  D.~Zagier,
  {\it The Eichler-Selberg trace formula on $SL_2(\mathbb{Z})$,} Appendix in S.~Lang {\it
  Introduction to modular forms} Springer-Verlag 1976, and {\it
  Correction} in {\it Modular functions of one variable VI,} Lect. in
  Math. {\bf 627}, Springer-Verlag 1976.

\bibitem{apostol:1976}
T.~M. Apostol, {\em Introduction to Analytic Number Theory}.
\newblock Springer-Verlag, 1976.

\bibitem{Skoruppa:1988}
  N.~P.~Skoruppa and D.~Zagier,
  {\it Jacobi forms and a certain space of modular forms},
  Invent.\ math.\  {\bf 94} (1988) 113.

\bibitem{Skoruppa:1989}
  N.~P.~Skoruppa and D.~Zagier,
  {\it A trace formula for Jacobi forms},
  J.\ reine\ angew.\ Math.\  {\bf 393} (1989) 168.

\bibitem{Skoruppa:2007}
  N.~P.~Skoruppa,
  {\it Memorandum on dimension formulas for spaces of Jacobi forms},
  arXiv:0711.0632 [math.NT].

\bibitem{Kramer:1991}
  J.~Kramer,
  {\it A geometrical approach to the theory of Jacobi forms},
  Comp. Math. {\bf 79} (1991) 1.

\bibitem{Gaberdiel:2008pr}
  M.~R.~Gaberdiel and C.~A.~Keller,
  {\it Modular differential equations and null vectors},
  arXiv:0804.0489 [hep-th].

\bibitem{Vafa:1994tf}
  C.~Vafa and E.~Witten,
  {\it A strong coupling test of S duality},
  Nucl.\ Phys.\  B {\bf 431} (1994) 3
  [arXiv:hep-th/9408074].

\bibitem{Gaiotto:2007cd}
  D.~Gaiotto and X.~Yin,
  {\it Examples of M5-brane elliptic genera},
  JHEP {\bf 0711} (2007) 004
  [arXiv:hep-th/0702012].

\bibitem{Denef:2007vg}
  F.~Denef and G.~W.~Moore,
  {\it Split states, entropy enigmas, holes and halos},
  arXiv:hep-th/0702146.

\bibitem{Maldacena:1999bp}
  J.~M.~Maldacena, G.~W.~Moore and A.~Strominger,
  {\it Counting BPS black holes in toroidal type II string theory},
  arXiv:hep-th/9903163.

\bibitem{Freed:1999vc}
  D.~S.~Freed and E.~Witten,
  {\it Anomalies in string theory with D-branes},
  arXiv:hep-th/9907189.

\bibitem{Montonen:1977sn}
  C.~Montonen and D.~I.~Olive,
  {\it Magnetic Monopoles As Gauge Particles?},
  Phys.\ Lett.\  B {\bf 72} (1977) 117.

\bibitem{Goddard:1976qe}
  P.~Goddard, J.~Nuyts and D.~I.~Olive,
  {\it Gauge Theories And Magnetic Charge},
  Nucl.\ Phys.\  B {\bf 125} (1977) 1.

\bibitem{Witten:1998wy}
  E.~Witten,
  {\it AdS/CFT correspondence and topological field theory},
  JHEP {\bf 9812} (1998) 012
  [arXiv:hep-th/9812012].

\bibitem{Maldacena:2001ss}
  J.~M.~Maldacena, G.~W.~Moore and N.~Seiberg,
  {\it D-brane charges in five-brane backgrounds},
  JHEP {\bf 0110} (2001) 005
  [arXiv:hep-th/0108152].

\bibitem{Moore:2004jv}
  G.~W.~Moore,
  {\it Anomalies, Gauss laws, and page charges in M-theory},
  Comptes Rendus Physique {\bf 6} (2005) 251
  [arXiv:hep-th/0409158].

\bibitem{ogg:1969}
A. Ogg, {\em Modular forms and Dirichlet series}.
\newblock W. A. Benjamin, Inc., 1969.

\bibitem{gottsche:1990}
  L.~G\"ottsche,
  {\it The Betti numbers of the Hilbert scheme of points on a smooth projective surface},
  Math. Ann. {\bf 286} (1990) 193.

\bibitem{Yoshioka:1994}
K.~Yoshioka,
{\it The Betti numbers of the moduli space of stable sheaves of rank 2 on $\mathbb{P}^2$},
J. reine. angew. Math. {\bf 453} (1994) 193.

\bibitem{Klyachko:1991}
 A.~A.~Klyachko,
 {\it Moduli of vector bundles and numbers of classes},
 Funct. Anal. and Appl. {\bf 25} (1991) 67. 

\bibitem{zagier:1975}
  D.~Zagier,
  {\it Nombres de classes et formes modulaires de poids 3/2, }
  C.R. \ Acad. \ Sc. \  Paris, {\bf 281} (1975) 883.

\bibitem{hirzebruch:1976}
  F.~Hirzebruch and D.~Zagier,
  {\it Intersection numbers of curves on Hilbert modular surfaces and modular
  forms of Nebentypus, }
  Inv.\ Math. \  {\bf 36} (1976) 57.

\bibitem{Minahan:1998vr}
  J.~A.~Minahan, D.~Nemeschansky, C.~Vafa and N.~P.~Warner,
  {\it E-strings and N = 4 topological Yang-Mills theories},
  Nucl.\ Phys.\  B {\bf 527} (1998) 581
  [arXiv:hep-th/9802168].






 



\end{thebibliography}
\end{document}